\documentclass[aps,prd,preprintnumbers,groupedaddress,nofootinbib,amssymb,notitlepage,eqsecnum]{revtex4-2}
\usepackage{here}
\usepackage[dvipdfmx]{graphicx}
\usepackage{amsmath,amsthm,amssymb}
\usepackage{bm}
\usepackage{color}

\usepackage{amsfonts}
\usepackage{dcolumn}
\usepackage{hyperref}
\allowdisplaybreaks[1]
\usepackage{stackengine}

\newcommand{\be}{\begin{equation}}  
\newcommand{\ee}{\end{equation}}
\newcommand{\ba}{\begin{eqnarray}}
\newcommand{\ea}{\end{eqnarray}}

\newcommand{\rd}{{\rm d}}

\newcommand{\bem}{\begin{bmatrix}}
\newcommand{\eem}{\end{bmatrix}}
\newcommand{\Mpl}{M_{\rm Pl}}

\newcommand{\mJ}{\mathcal{J}}

\newcommand{\mG}{\mathcal{G}}

\allowdisplaybreaks

\begin{document}

\preprint{YITP-23-38, WUCG-23-03}
\title{Coupled vector Gauss-Bonnet theories and hairy black holes}

\author{Katsuki Aoki$^{1}$ and Shinji Tsujikawa$^{2}$}

\affiliation{
$^1$Center for Gravitational Physics and Quantum Information, 
Yukawa Institute for Theoretical Physics, Kyoto University, 
606-8502, Kyoto, Japan\\
$^2$Department of Physics, Waseda University, 3-4-1 Okubo, 
Shinjuku, Tokyo 169-8555, Japan}

\begin{abstract}

We study vector-tensor theories in which a 4-dimensional 
vector field $A_{\mu}$ is coupled to a vector quantity 
${\cal J}^{\mu}$, which is expressed in terms of $A_{\mu}$ 
and a metric tensor $g_{\mu \nu}$. 
The divergence of ${\cal J}^{\mu}$ is equivalent to 
a Gauss-Bonnet (GB) term. We show that an interacting Lagrangian 
of the form $f(X)A_{\mu}{\cal J}^{\mu}$, where $f$ is an 
arbitrary function of $X=-(1/2)A_{\mu}A^{\mu}$, belongs to 
a scheme of beyond generalized Proca theories. 
For $f(X)=\alpha={\rm constant}$, this interacting Lagrangian 
reduces to a particular class of generalized Proca theories. 
We apply the latter coupling to a static and spherically 
symmetric vacuum configuration by incorporating 
the Einstein-Hilbert term, Maxwell scalar, and 
vector mass term $\eta X$ ($\eta$ is a constant). 
Under an expansion of the small coupling constant $\alpha$ 
with $\eta \neq 0$, we derive hairy black hole solutions 
endowed with nonvanishing temporal and radial vector 
field profiles. The asymptotic properties of solutions 
around the horizon and at spatial infinity are different from 
those of hairy black holes present in scalar-GB theories. 
We also show that black hole solutions without the vector mass term, 
i.e., $\eta=0$, are prone to ghost instability of odd-parity perturbations.

\end{abstract}

\date{\today}


\maketitle

\section{Introduction}
\label{introsec}

General Relativity (GR) has been well tested in solar-system 
experiments \cite{Will:2014kxa}
and submillimeter laboratory tests of gravity \cite{Hoyle:2000cv,Adelberger:2003zx}. 
If we go beyond the solar-system scales, however, there are several 
unsolved problems such as the origins of dark energy 
and dark matter \cite{Copeland:2006wr,Bertone:2004pz}. 
At very high energy close to the Planck scale, we also believe that GR 
should be replaced by a more fundamental theory with 
an ultraviolet completion. In such extreme gravity regimes,
it is expected that some higher-order curvature corrections to  
the Einstein-Hilbert action come into play. 
These curvature corrections can potentially modify the physics 
of highly compact objects like black holes (BHs) 
and neutron stars.
After the dawn of gravitational wave astronomy \cite{LIGOScientific:2016aoc}, 
one can now probe signatures for the possible 
deviation from GR in strong gravity regimes \cite{Berti:2015itd,Barack:2018yly,Berti:2018cxi}.

For the construction of healthy gravitational theories, 
it is desirable to keep the field equations of motion 
up to second order in the metric tensor $g_{\mu \nu}$. 
In this case, one can avoid so-called Ostrogradski 
instability \cite{Ostrogradsky:1850fid,Woodard:2015zca} 
arising from higher-order derivative terms.  
Using a general class of Lagrangians containing polynomial 
functions of Riemann curvature tensors, 
Lanczos \cite{Lanczos:1938sf} and Lovelock \cite{Lovelock:1971yv}
constructed gravitational theories with second-order field 
equations of motion. In the 4-dimensional spacetime,  
the field equations uniquely reduce to those in GR. 
In spacetime dimensions $D$ higher than 4, Lanczos and Lovelock 
theories differ from GR and have richer structures. 
In particular, there is a quadratic-order curvature scalar known 
as a Gauss-Bonnet (GB) term modifying the 
spacetime dynamics in $D>4$ dimensions \cite{Stelle:1977ry}.

When $D=4$, the GB term is a topological surface term which 
does not contribute to the field equations of motion.
If there is a scalar field $\phi$ coupled to the GB term $\mG$ of the 
form $\mu (\phi) \mG$, where $\mu$ is a function of $\phi$, 
the 4-dimensional spacetime dynamics is modified by the scalar-GB coupling. In string theory, for example, the low energy 
effective action contains a coupling between the dilaton 
field $\phi$ and $\mG$ of the form $e^{-\lambda \phi}\mG$ \cite{Zwiebach:1985uq,Gross:1986mw,Metsaev:1987zx}.  
If we apply the scalar-GB coupling $\mu (\phi) \mG$ to 
a spherically symmetric configuration, it is known that 
there are hairy BH and neutron star solutions with 
nontrivial scalar field profiles \cite{Kanti:1995vq,Torii:1996yi,Chen:2006ge,Guo:2008hf,Pani:2009wy,Kleihaus:2011tg,Sotiriou:2013qea,Sotiriou:2014pfa,Ayzenberg:2014aka,Maselli:2015tta,Kleihaus:2015aje,Doneva:2017bvd,Silva:2017uqg,Antoniou:2017acq,Minamitsuji:2018xde,Silva:2018qhn,Langlois:2022eta,Minamitsuji:2022vbi,Minamitsuji:2022mlv,Minamitsuji:2022tze}.
The role of the same scalar-GB coupling in cosmology has 
been also extensively studied in the literature \cite{Antoniadis:1993jc,Gasperini:1996fu,Kawai:1998ab,Cartier:1999vk,Cartier:2001is,Tsujikawa:2002qc,Toporensky:2002ta,Amendola:2005cr,Nojiri:2005vv,Calcagni:2005im,Calcagni:2006ye,Koivisto:2006xf,Koivisto:2006ai,Tsujikawa:2006ph,Guo:2006ct,Amendola:2007ni,Satoh:2008ck,Guo:2009uk,Kanti:2015pda,Hikmawan:2015rze,Kawai:2017kqt,Yi:2018gse,Barton:2021wfj,Kawai:2021edk,Zhang:2021rqs,Kawai:2021bye,Kawaguchi:2022nku}. 
We note that the extension to more general scalar-GB couplings  
$f(\phi,\mG)$ containing nonlinear functions of $\mG$ leads 
to instabilities of scalar perturbations associated with the nonlinear 
GB term during decelerating cosmological epochs \cite{Tsujikawa:2022aar} 
(see also Refs.~\cite{Carroll:2004de,Chiba:2005nz,DeFelice:2006pg,DeFelice:2009rw,DeFelice:2010sh,DeFelice:2009wp,DeFelice:2009ak,DeFelice:2009rw}).

The linear scalar-GB coupling $\phi \mG$ has a peculiar property in four dimensions. The action $\int {\rm d}^4x \sqrt{-g}\,\phi\mG$ is invariant under the global shift of the scalar field $\phi \to \phi + c$ with $c$ being constant,  as the integral $c \int {\rm d}^4x \sqrt{-g}\,\mG$ is a boundary term. 
When a theory enjoys the global symmetry, one may promote it to a local symmetry by introducing a gauge field $A_{\mu}$ and replacing derivatives with covariant derivatives. When the shift symmetry is localized, the scalar field $\phi$ corresponds to a gauge mode and can be eliminated by fixing the gauge. 
Then, the resulting theory is a vector-tensor theory having three dynamical degrees of freedom (DOFs) on top of the DOFs of spacetime metric $g_{\mu\nu}$ (cf., Refs.~\cite{Cheng:2006us,Mukohyama:2006mm,Aoki:2021wew}). 
In this paper, we shall apply this idea to the linear scalar-GB coupling and find a vector-tensor theory analogous to the scalar-GB theory.

In practice, the above procedure requires to find a vector quantity 
${\cal J}^{\mu}$ whose divergence is equivalent to $\mG$, i.e., 
$\nabla_{\mu}{\cal J}^{\mu}=\mG$, by which the linear scalar-GB coupling can be recast in the form $-\int {\rm d}^4x \sqrt{-g}\,\nabla_{\mu} \phi {\cal J}^{\mu}$ via integration by parts. 
The integral ${\cal J}^{\mu}$ may not be unique since it is defined only through the differential equation $\nabla_{\mu}{\cal J}^{\mu}=\mG$. 
One form of ${\cal J}^{\mu}$, which is expressed in terms of 
a scalar field and Riemann tensor, is found in Ref.~\cite{Colleaux:2019ckh}.
On using this expression of ${\cal J}^{\mu}={\cal J}^{\mu}[\phi,g]$ 
and the property $\mG= \nabla_{\mu}{\cal J}^{\mu}$, 
it is possible to prove that the scalar-GB coupling $\mu (\phi)\mG$ 
belongs to a subclass of Horndeski theories \cite{Horndeski} 
after the integration by parts \cite{Langlois:2022eta}.
We note that the equivalence between the scalar-GB coupling and 
Horndeski theories was originally shown in Ref.~\cite{KYY} 
by taking the approach of field equations of motion. 

We will find an alternative expression of ${\cal J}^{\mu}$ by using 
a vector field $A_{\mu}$, where ${\cal J}^{\mu}={\cal J}^{\mu}[A,g]$ 
satisfies the same relation $\nabla_{\mu}{\cal J}^{\mu}=\mG$.
As a candidate for a Lorentz-invariant scalar characterizing 
the coupling between $A_{\mu}$ and the integrated GB term 
in vector-tensor theories, we propose the Lagrangian 
$A_{\mu}{\cal J}^{\mu}$.
We will show that the interacting Lagrangian 
$A_{\mu}{\cal J}^{\mu}$ is equivalent to a subclass of generalized 
Proca (GP) theories with second-order field equations of motion \cite{Heisenberg:2014rta,Tasinato:2014eka,Allys:2015sht,BeltranJimenez:2016rff,GallegoCadavid:2019zke}, and by construction, it reduces to a linear scalar-GB coupling in a certain limit.
We will also extend the analysis to a more general Lagrangian 
$f(X)A_{\mu}{\cal J}^{\mu}$, where $f$ is an arbitrary function 
of $X=-(1/2)A_{\mu} A^{\mu}$. 
In this case, the resulting vector-tensor theory is shown to be 
equivalent to a class of beyond GP theories originally proposed in 
Ref.~\cite{Heisenberg:2016eld} (see also Ref.~\cite{Kimura:2016rzw}).
Since beyond GP theories correspond to a healthy extension 
of GP theories with the same dynamical DOFs, 
we are now able to construct healthy theories of 
a vector field coupled to the integrated GB term. 

We will also apply the interacting Lagrangian 
$\alpha A_{\mu}{\cal J}^{\mu}$ 
($\alpha$ is a coupling constant) 
to the search for hairy BH solutions 
on a static and spherically symmetric background. 
For this purpose, we also take into account 
the Einstein-Hilbert term, Maxwell term, and 
vector mass term $\eta X$, 
where $\eta$ is a constant.
The coupling $\alpha A_{\mu}{\cal J}^{\mu}$ is equivalent to 
a Lagrangian of the quintic-order coupling function  
$G_5(X)=4\alpha \ln |X|$ in GP theories.
In Refs.~\cite{Heisenberg:2017xda,Heisenberg:2017hwb}, it was shown 
that there are no hairy BH solutions with regular vector field 
profiles for the positive power-law 
quintic functions $G_5 \propto X^n$ with $n \geq 1$. 
However, we will show that this is not the case for 
$G_5(X)=4\alpha \ln |X|$. 
Under an expansion of the small coupling constant $\alpha$, 
we derive solutions to the temporal and radial vector components 
around the BH horizon and at spatial infinity. 
Numerically it is challenging to perform accurate integrations   
due to the existence of a rapidly growing mode arising from
the mass term $\eta X$, but we are able to find out solutions 
that mimic the asymptotic behavior in some large-distance 
regions. In comparison to hairy BHs present for the linear scalar-GB 
coupling, the behavior of hairy BH solutions for $\eta \neq 0$ 
is different both around the horizon and at large distances.
We will also show that BH solutions for $\eta=0$ suffer from ghost 
instability of odd-parity perturbations. 

This paper is organized as follows. 
In Sec.~\ref{scasec}, we first review the correspondence between 
the scalar-GB coupling $\mu(\phi)\mG$ and the Horndeski Lagrangian. 
We then introduce a vector field ${\cal J}^{\mu}$ whose 
divergence $\nabla_{\mu}{\cal J}^{\mu}$ is equivalent to the GB term 
and show that the Lagrangian $f(X) A_{\mu}{\cal J}^{\mu}$ belongs to 
a subclass of beyond GP theories. 
In Sec.~\ref{BHsec}, we study static and spherically symmetric 
BH solutions for the coupling $\alpha A_{\mu}{\cal J}^{\mu}$ 
and derive perturbative solutions to $A_{\mu}$ with respect to 
the small coupling $\alpha$. 
We also numerically confirm the existence of vector field profiles 
connecting two asymptotic regimes and analytically estimate 
corrections to the gravitational potentials 
arising from the vector-GB couplings.
Sec.~\ref{consec} is devoted to conclusions.

\section{Coupled vector Gauss-Bonnet theories}
\label{scasec}

The GB curvature invariant is a specific combination 
of Lanczos \cite{Lanczos:1938sf} and 
Lovelock \cite{Lovelock:1971yv} scalars. 
In 4-dimensional spacetime, the GB term 
is given by \cite{Padmanabhan:2013xyr,Fernandes:2022zrq}
\be
\mG=\frac{1}{4}\delta^{\mu\nu\rho\sigma}
_{\alpha\beta\gamma\delta}
R^{\alpha\beta}{}_{\mu\nu}R^{\gamma\delta}{}_{\rho\sigma}\,,
\label{Gdef}
\ee
where $\delta^{\mu_1 \cdots \mu_k}_{\nu_1 \cdots \nu_k}=
k! \delta^{[\mu_1}_{\nu_1} \cdots \delta^{\mu_k]}_{\nu_k}$ is 
the generalized Kronecker delta 
and $R^{\alpha\beta}{}_{\mu\nu}$ is the Riemann tensor. 
More explicitly, Eq.~(\ref{Gdef}) can be expressed as
\be
\mG = R^2-4R_{\mu\nu}R^{\mu\nu}
+R_{\mu\nu\rho\sigma}R^{\mu\nu\rho\sigma}\,,
\ee
where $R$ is the scalar curvature and $R_{\mu \nu}$ 
is the Ricci tensor. In 4 dimensions, the GB term is 
a total derivative and does not contribute to the equations 
of motion while the 4-dimensional spacetime 
dynamics is modified in the presence of a scalar or vector field 
coupled to $\mG$ or its associated vector.

\subsection{Scalar field coupled to the GB term}

Let us first briefly revisit the case in which there is 
a scalar field $\phi$ coupled to the GB term of 
the form $\mu (\phi) \mG$. 
Because of the antisymmetric property of 
$\delta^{\mu_1 \cdots \mu_k}_{\nu_1 \cdots \nu_k}$, 
the field equations of motion following from 
the coupling $\mu (\phi) \mG$ are of second order 
in the metric tensor $g_{\mu \nu}$ and the scalar field $\phi$. 
On using the Riemann tensor and covariant derivatives of $\phi$, 
the GB term can be expressed in the form \cite{Colleaux:2019ckh,Langlois:2022eta}
\be
\mG=\delta^{\mu \nu \rho \sigma}_{\alpha \beta \gamma \delta}
\nabla^{\delta} \left[ 
\frac{\nabla^{\gamma} \nabla_{\rho} \phi \nabla_{\sigma}\phi}{X_s} 
\left( {R^{\alpha \beta}}_{\mu \nu}
-\frac{2}{3X_s} \nabla^\alpha \nabla_{\mu}\phi 
\nabla^\beta \nabla_{\nu} \phi \right)
\right]\,, 
\label{mG}
\ee
where $\nabla^\delta$ is a covariant derivative operator and 
$X_s=-(1/2)\nabla_{\mu}\phi \nabla^{\mu}\phi$. 
Substituting this expression of $\mG$ into the action 
\be
{\cal S}_{\rm sGB}=\int \rd^4 x \sqrt{-g}\,\mu(\phi)\mG\,,
\label{SsGB}
\ee
and integrating (\ref{SsGB}) by parts, it follows that 
\be
{\cal S}_{\rm sGB}=-\int \rd^4 x \sqrt{-g}\,\mu_{,\phi}(\phi)
\delta^{\mu \nu \rho \sigma}_{\alpha \beta \gamma \delta}
\frac{1}{X_s} \nabla^{\gamma} \nabla_{\rho} \phi 
\nabla_{\sigma}\phi \nabla^{\delta} \phi
\left( {R^{\alpha \beta}}_{\mu \nu}
-\frac{2}{3X_s} \nabla^\alpha \nabla_{\mu}\phi 
\nabla^\beta \nabla_{\nu} \phi \right)\,,
\label{SGB}
\ee
where $g$ is the determinant of $g_{\mu \nu}$, and 
we use the notations $F_{,\phi} \equiv \partial F/\partial \phi$ 
and $F_{,X_s} \equiv \partial F/\partial X_s$ for any arbitrary function $F$. 
We will expand the generalized Kronecker delta, integrate 
the action (\ref{SGB}) by parts, and exploit the relation 
$[\nabla_{\mu}, \nabla_{\nu}]\nabla^{\alpha}\phi
={R^{\alpha}}_{\lambda \mu \nu}\nabla^{\lambda}\phi$ 
to eliminate contractions of the Riemann tensors. 
Up to boundary terms, the action (\ref{SGB}) 
is equivalent to \cite{Langlois:2022eta}
\ba
\hspace{-0.7cm}
{\cal S}_{\rm sGB}
&=& 
\int {\rm d}^4 x \sqrt{-g}\,
\biggl[ G_{2s}-G_{3s}\square\phi 
+G_{4s}R +G_{4s,X_s} \left\{ (\square \phi)^{2}
-(\nabla_{\mu}\nabla_{\nu} \phi)
(\nabla^{\mu}\nabla^{\nu} \phi) \right\} \nonumber \\
\hspace{-0.7cm}
& &
+G_{5s} G_{\mu \nu} \nabla^{\mu}\nabla^{\nu} \phi
-\frac{1}{6}G_{5s,X_s}
\left\{ (\square \phi )^{3}-3(\square \phi)\,
(\nabla_{\mu}\nabla_{\nu} \phi)
(\nabla^{\mu}\nabla^{\nu} \phi)
+2(\nabla^{\mu}\nabla_{\alpha} \phi)
(\nabla^{\alpha}\nabla_{\beta} \phi)
(\nabla^{\beta}\nabla_{\mu} \phi) \right\} 
\biggr]\,,
\label{SHo}
\ea
where $G_{\mu \nu}=R_{\mu \nu}-(1/2) g_{\mu \nu}R$ is 
the Einstein tensor, and 
\ba
& &
G_{2s}=-8 \mu_{,\phi \phi \phi \phi}(\phi) X_s^2 (3-\ln |X_s|)\,,\qquad 
G_{3s}=4 \mu_{,\phi \phi \phi}(\phi) X_s (7-3\ln |X_s|)\,,\nonumber \\
& &
G_{4s}=4 \mu_{,\phi \phi}(\phi) X_s (2-\ln |X_s|)\,,\qquad
G_{5s}=-4 \mu_{,\phi}(\phi) \ln |X_s|\,.
\label{Ggauss}
\ea
The action (\ref{SHo}) belongs to a subclass of scalar 
Horndeski theories \cite{Horndeski} with second-order Euler equations 
of motion. Originally, the equivalence of scalar-GB theories with 
Horndeski theories given by the coupling functions (\ref{Ggauss}) 
was shown in Ref.~\cite{KYY} by using 
the field equations of motion.
In Ref.~\cite{Langlois:2022eta}, the same equivalence was proven 
at the level of the action (as explained above). 

For the linear scalar-GB coupling $\mu(\phi)=-\alpha \phi$, where 
$\alpha$ is a constant, we have $G_{5s}=4\alpha \ln |X_s|$ and
$G_{2s}=G_{3s}=G_{4s}=0$. This falls into a subclass of 
shift-symmetric Horndeski theories where the field equations 
of motion are invariant under the shift $\phi \to \phi+c$.
In the original form of the GB coupling $\phi \mG$, the action is quasi-invariant under the shift $\phi \to \phi+c$, i.e., invariant up to a total derivative, while the Lagrangian in the Horndeski form is manifestly invariant under the shift.
For $\mu(\phi)$ containing nonlinear functions of $\phi$, 
we generally have the $\phi$ dependence
in $G_{2s}, G_{3s}, G_{4s}, G_{5s}$. 
As we mentioned in Introduction, there are BH and neutron star 
solutions endowed with scalar hairs for such scalar-GB couplings.

\subsection{Vector field coupled to the integrated GB term}

If we want to construct theories in which a vector field $A_{\mu}$ 
is coupled to the GB term in some way, we need to construct 
a Lorentz-invariant scalar appearing in the Lagrangian. For instance, one may consider the coupling $A^{\mu}\nabla_{\mu} \mG$. However, the equations of motion associated with this coupling contain derivatives higher than second order and hence such theories are generally prone to Ostrogradski instability. 
Another possible coupling would be $\mu_{\rm v}(X) \mG$
where 
\be
X \equiv -\frac{1}{2} A_{\mu} A^{\mu}\,.
\ee
Again, this interaction may summon a ghostly DOF in the longitudinal sector 
of $A_{\mu}$ which can be understood by taking the decoupling of the longitudinal and transverse DOFs. The longitudinal mode becomes manifest by introducing the St\"{u}ckelberg field according to the replacement $A_{\mu} \to g_{\rm v} A_{\mu}+\nabla_{\mu}\phi$. 
Then, the decoupling limit $g_{\rm v} \to 0$ gives $X \to X_s$. 
The interacting Lagrangian $\mu_{\rm v}(X) \mG$ reduces to a coupling between the GB term and the derivative of $\phi$, not the scalar field itself, which should yield equations of motion with derivatives higher than second order.

As we already explained, the linear coupling $-\phi \mathcal{G}$ has a global shift symmetry and the vector-tensor theory can be obtained by localizing this global symmetry. This requires finding a vector field $\mJ^{\mu}$ whose divergence agrees with the GB term, $\nabla_{\mu}\mJ^{\mu} = \mG$. After integration by parts, the coupling $-\phi \mathcal{G}$ becomes $\nabla_{\mu}\phi \mJ^{\mu}[\phi, g]$. As shown in Eq.~\eqref{SGB}, the action contains the derivatives of $\phi$ but not the field itself, for the linear coupling $\mu_{,\phi}=-\alpha$. The global shift symmetry can be localized by the replacement 
$\nabla_{\mu} \phi \to \nabla_{\mu}\phi + g_{\rm v} A_{\mu}$ with the help of the vector field $A_{\mu}$. The symmetry transformation is now $\phi \to \phi + \chi(x), A_{\mu} \to A_{\mu}-\nabla_{\mu} \chi(x)/g_{\rm v}$. The scalar field $\phi$ can be eliminated by setting the unitary gauge $\phi=0$. All in all, what we need is the replacement 
$\nabla_{\mu}\phi \to A_{\mu}$, where the gauge coupling $g_{\rm v}$ 
is absorbed into the definition of $A_{\mu}$. 
In this way, we can obtain a vector-tensor theory coupled 
to the integrated GB term.

However, there is an ambiguity in the above procedure. The second derivative 
$\nabla_{\mu}\nabla_{\nu}\phi$ is symmetric 
in its indices while the replaced quantity $\nabla_{\mu}A_{\nu}$ is not symmetric. We resolve this ambiguity by imposing the condition  $\nabla_{\mu}\mJ^{\mu} = \mG$ even after the replacement 
$\nabla_{\mu}\phi \to A_{\mu}$. 
The expression of $\mathcal{J}^{\mu}$ consistent with such 
requirement is given by 
\begin{align}
\mathcal{J}^{\mu} \equiv \delta^{\mu\nu\rho\sigma}_{\alpha\beta\gamma\delta}
\left[ \frac{A^{\alpha} \nabla_{\nu}A^{\beta}}{X} 
\left( R^{\gamma\delta}{}_{\rho\sigma}-\frac{2}{3X}\nabla_{\rho}
A^{\gamma}\nabla_{\sigma}A^{\delta} \right) \right]\,.
\label{Jmu}
\end{align}
In the following, we will show that this vector field satisfies 
the relation $\nabla_{\mu}\mJ^{\mu} = \mG$. 
In doing so, we use the relation 
$\nabla_{[\mu}R^{\alpha\beta}{}_{\nu\rho]}=0$ and 
the antisymmetric property of the generalized Kronecker delta. 
We also exploit the following equality
\be
\delta^{\mu_1 \mu_2 \mu_3 \mu_4 }_{\nu_1 \nu_2 \nu_3 \nu_4} 
\nabla_{\mu_1}\left( \frac{A^{\nu_1 }}{X^2} \right) \nabla_{\mu_2}A^{\nu_2} 
\nabla_{\mu_3}A^{\nu_3}\nabla_{\mu_4}A^{\nu_4} 
=
-\frac{1}{2X^3} \delta^{\mu_1 \mu_2 \mu_3 \mu_4 \mu_5}_{\nu_1 \nu_2 \nu_3 \nu_4 \nu_5} 
A_{\mu_1}A^{\nu_1}  \nabla_{\mu_2}A^{\nu_2} \nabla_{\mu_3}A^{\nu_3}
\nabla_{\mu_4}A^{\nu_4} \nabla_{\mu_5}A^{\nu_5}=0\,,
\label{delta5}
\ee
together with the expansion of 
$\delta^{\mu_1\cdots \mu_d}_{\nu_1 \cdots \nu_d}$
in $d$ dimensions:
\be
\delta^{\mu_1\cdots \mu_d}_{\nu_1 \cdots \nu_d}
=\sum_{k=1}^d (-1)^{d+k} 
\delta^{\mu_d}_{\nu_k}
\delta^{\mu_1 \cdots \mu_k \cdots \mu_{d-1}}
_{\nu_1 \cdots \bar{\nu}_k \cdots \nu_{d}}\,,
\label{delexpan}
\ee
where $\bar{\nu}_k$ means that this index is omitted. 
Notice that the 5-dimensional generalized Kronecker delta 
$\delta^{\mu_1 \mu_2 \mu_3 \mu_4 \mu_5}_{\nu_1 \nu_2 \nu_3 \nu_4 \nu_5}$ 
vanishes in 4-dimensional spacetime, whose property was used 
in the second equality of Eq.~\eqref{delta5}.
Then, it follows that 
\be
\nabla_{\mu}\mathcal{J}^{\mu}=
{\cal F}_1+{\cal F}_2+{\cal F}_3\,,
\label{Jmu1}
\ee
where 
\ba
{\cal F}_1 &=& \delta^{\mu\nu\rho\sigma}_{\alpha\beta\gamma\delta}
\frac{A^{\alpha}}{X} \nabla_{\mu} \nabla_{\nu}A^{\beta}R^{\gamma\delta}{}_{\rho\sigma}\,,
\label{F1} \\ 
{\cal F}_2 &=& \delta^{\mu\nu\rho\sigma}_{\alpha\beta\gamma\delta}
\nabla_{\mu} \left( \frac{A^{\alpha}}{X} \right) \nabla_{\nu}A^{\beta}R^{\gamma\delta}{}_{\rho\sigma}\,,\\ 
{\cal F}_3 &=&-\frac{2}{3}\delta^{\mu\nu\rho\sigma}_{\alpha\beta\gamma\delta}
\frac{A^{\alpha}}{X^2} \nabla_{\mu} \left( \nabla_{\nu}A^{\beta}
\nabla_{\rho}A^{\gamma}\nabla_{\sigma}A^{\delta} \right)\,.
\label{F3}
\ea
Since $\nabla_{\mu} \nabla_{\nu}A^{\beta}
-\nabla_{\nu} \nabla_{\mu}A^{\beta}=
{R^{\beta}}_{\lambda \mu \nu}A^{\lambda}$, 
Eq.~(\ref{F1}) reduces to 
\be
{\cal F}_1=\frac{1}{2}
\delta^{\mu\nu\rho\sigma}_{\alpha\beta\gamma\delta}
\frac{A^{\alpha}A^{\lambda}}{X} 
{R^{\beta}}_{\lambda\mu \nu}
R^{\gamma\delta}{}_{\rho\sigma}
=\frac{1}{4}
\delta^{\mu\nu\rho\sigma}_{\alpha\beta\gamma\delta}
{R^{\alpha \beta}}_{\mu \nu}
R^{\gamma\delta}{}_{\rho\sigma}=\mG\,,
\label{F1re}
\ee
where, in the second equality, we used the relation 
\be
\delta^{\mu_1 \mu_2 \mu_3 \mu_4 }_{\nu_1 \nu_2 \nu_3 \nu_4}\left(\frac{2A^{\nu_1}A^{\lambda}}{X}R^{\nu_2}{}_{\lambda \mu_1 \mu_2} R^{\nu_3 \nu_4}{}_{\mu_3 \mu_4}-R^{\nu_1\nu_2}{}_{\mu_1\mu_2} R^{\nu_3 \nu_4}{}_{\mu_3 \mu_4}  \right)
=\frac{1}{2X}  \delta^{\mu_1 \mu_2 \mu_3 \mu_4 \mu_5}_{\nu_1 \nu_2 \nu_3 \nu_4 \nu_5} A_{\mu_1}A^{\nu_1}  R^{\nu_2\nu_3}{}_{\mu_2\mu_3}  R^{\nu_4 \nu_5}{}_{\mu_4 \mu_5}=0\,.
\ee
For the computation of ${\cal F}_2$, we exploit the following property
\ba 
& &
\delta^{\mu_1 \mu_2 \mu_3 \mu_4 }_{\nu_1 \nu_2 \nu_3 \nu_4} \left[\nabla_{\mu_1}\left( \frac{A^{\nu_1 }}{X} \right) \nabla_{\mu_2}A^{\nu_2} R^{\nu_3\nu_4}{}_{\mu_3\mu_4 }- \frac{A^{\nu_1}A^{\alpha}}{X^2} R^{\nu_2}{}_{\alpha \mu_1 \mu_2} \nabla_{\mu_3}A^{\nu_3}\nabla_{\mu_4}A^{\nu_4} \right]
\nonumber \\
& &
=-\frac{1}{2X^2} \delta^{\mu_1 \mu_2 \mu_3 \mu_4 \mu_5}_{\nu_1 \nu_2 \nu_3 \nu_4 \nu_5} A_{\mu_1}A^{\nu_1}  R^{\nu_2\nu_3}{}_{\mu_2\mu_3} \nabla_{\mu_4}A^{\nu_4} \nabla_{\mu_5}A^{\nu_5}=0\,.
\ea
Then, we have 
\be
{\cal F}_2=
\delta^{\mu\nu\rho\sigma}_{\alpha\beta\gamma\delta}
\frac{A^{\alpha} A^{\lambda}}{X^2}
{R^{\beta}}_{\lambda \mu \nu} 
\nabla_{\rho} A^{\gamma} \nabla_{\sigma} A^{\delta}\,.
\ee
Expanding the covariant derivative $\nabla_{\mu}$ in ${\cal F}_3$ and 
using the commutation relation of $\nabla_{\mu}\nabla_{\nu}A^{\beta}$,  
we obtain 
\be
{\cal F}_3=-\delta^{\mu\nu\rho\sigma}_{\alpha\beta\gamma\delta}
\frac{A^{\alpha} A^{\lambda}}{X^2}
{R^{\beta}}_{\lambda \mu \nu} 
\nabla_{\rho} A^{\gamma} \nabla_{\sigma} A^{\delta}
=-{\cal F}_2\,.
\label{F3re}
\ee
On using Eqs.~(\ref{F1re}) and (\ref{F3re}), Eq.~(\ref{Jmu1}) yields
\be
\nabla_{\mu} \mathcal{J}^{\mu}=\mG\,.
\label{Jmu2}
\ee
Thus, the divergence of $\mathcal{J}^{\mu}$ 
is equivalent to the GB term. 

We consider a scalar quantity $A_{\mu}\mathcal{J}^{\mu}$ 
as a candidate for the ghost-free Lagrangian of the vector 
field $A_{\mu}$ coupled to the integrated GB term.
As a generalization, we also multiply an arbitrary function $f(X)$ of $X$ with 
the scalar product $A_{\mu}\mathcal{J}^{\mu}$. 
The interaction part of the action in such theories is given by 
\begin{align}
{\cal S}_{\rm vGB}=\int {\rm d}^4 x \sqrt{-g}\,
f(X) A_{\mu}\mathcal{J}^{\mu}\,,
\end{align}
which is composed of two parts:
\ba
{\cal S}_{{\rm vGB}1} &=& \int {\rm d}^4 x \sqrt{-g}\,
\delta^{\mu\nu\rho\sigma}_{\alpha\beta\gamma\delta}
\frac{f(X)}{X} A_{\mu} A^{\alpha} 
\nabla_{\nu}A^{\beta} {R^{\gamma \delta}}_{\rho \sigma}\,,
\label{SGB1} \\
{\cal S}_{{\rm vGB}2} &=& -\int {\rm d}^4 x \sqrt{-g}\,
\delta^{\mu\nu\rho\sigma}_{\alpha\beta\gamma\delta}
\frac{2f(X)}{3X^2} A_{\mu}A^{\alpha} 
\nabla_{\nu}A^{\beta}
\nabla_{\rho}A^{\gamma} \nabla_{\sigma}A^{\delta}\,.
\label{SGB2} 
\ea
On using the properties $\nabla_{\mu}X=-A_{\nu}\nabla_{\mu}A^{\nu}$ 
and $\nabla_{\mu} \ln X=-A_{\nu}\nabla_{\mu}A^{\nu}/X$, we find that 
Eqs.~(\ref{SGB1}) and (\ref{SGB2}) reduce, respectively, to 
\ba
\hspace{-1cm}
{\cal S}_{{\rm vGB}1} &=& \int {\rm d}^4 x \sqrt{-g}\, 
\biggl[ G_5(X) G_{\mu \nu} \nabla^{\mu} A^{\nu}
\nonumber\\
\hspace{-1cm}
& &\qquad \qquad \qquad
-\frac{4f(X)}{X} \left(A^{\rho} \nabla_{\sigma}A^{\sigma}
-A^{\sigma} \nabla_{\sigma} A^{\rho} \right) \nabla_{\nu}
\nabla_{\rho}A^{\nu}
-\frac{4f(X)}{X} \left(A^{\sigma} \nabla_{\sigma}A^{\nu}
-A^{\nu} \nabla_{\sigma} A^{\sigma} \right) \nabla_{\nu}
\nabla_{\rho}A^{\rho} 
\biggr]\,,\label{SGB1a} \\
\hspace{-1cm}
{\cal S}_{{\rm vGB}2} &=& \int {\rm d}^4 x \sqrt{-g}\, 
\biggl[ -\frac{2f(X)}{3X} \delta^{\mu\nu\rho}_{\alpha\beta\gamma}
\nabla_{\mu}A^{\alpha}\nabla_{\nu}A^{\beta} \nabla_{\rho}A^{\gamma} 
-3f_5(X) \delta^{\mu \nu \rho}_{\alpha \beta \gamma} 
A^{\alpha}A_{\lambda} \nabla_{\mu}A^{\lambda} \nabla_{\nu}A^{\beta} 
\nabla_{\rho}A^{\gamma}
\nonumber\\
\hspace{-1cm}
& &\qquad \qquad \qquad+\frac{4f(X)}{X} \left(A^{\rho} \nabla_{\sigma}A^{\sigma}
-A^{\sigma} \nabla_{\sigma} A^{\rho} \right) \nabla_{\nu}
\nabla_{\rho}A^{\nu}
+\frac{4f(X)}{X} \left(A^{\sigma} \nabla_{\sigma}A^{\nu}
-A^{\nu} \nabla_{\sigma} A^{\sigma} \right) \nabla_{\nu}
\nabla_{\rho}A^{\rho} 
\biggr]\,,\label{SGB2a}
\ea
where
\be
G_5(X) \equiv 4\int {\rm d}\tilde{X} \frac{f(\tilde{X})}{\tilde{X}}+8f(X)\,,\qquad 
f_5(X) \equiv -\frac{2f_{,X}}{3X}\,.
\label{G5}
\ee
Taking the sum of ${\cal S}_{{\rm vGB}1}$ and ${\cal S}_{{\rm vGB}2}$, 
terms on the second lines of Eqs.~(\ref{SGB1a}) and (\ref{SGB2a}) cancel each other. 
On using the property
\be
f_5 \delta^{\mu\nu\rho\sigma}_{\alpha\beta\gamma\delta}A_{\mu}A^{\alpha}
\nabla_{\nu}A^{\beta} \nabla_{\rho}A^{\gamma} \nabla_{\sigma}A^{\delta}
=\frac{4}{3} f_{,X}\delta^{\mu\nu\rho}_{\alpha\beta\gamma}\nabla_{\mu}
A^{\alpha}\nabla_{\nu}A^{\beta} \nabla_{\rho}A^{\gamma} 
-3f_5\delta^{\mu \nu \rho}_{\alpha \beta \gamma} 
A^{\alpha}A_{\lambda} \nabla_{\mu}A^{\lambda} \nabla_{\nu}A^{\beta} 
\nabla_{\rho}A^{\gamma}\,,
\ee
we obtain the reduced action 
\be
{\cal S}_{\rm vGB}=\int {\rm d}^4 x \sqrt{-g} \left[ G_5(X) G_{\mu\nu}\nabla^{\mu}A^{\nu} 
-\frac{1}{6}G_{5,X}(X)\delta^{\mu\nu\rho}_{\alpha\beta\gamma}\nabla_{\mu}A^{\alpha}\nabla_{\nu}A^{\beta} \nabla_{\rho}A^{\gamma} 
+ f_5(X) \delta^{\mu\nu\rho\sigma}_{\alpha\beta\gamma\delta}A_{\mu}A^{\alpha}
\nabla_{\nu}A^{\beta} \nabla_{\rho}A^{\gamma} \nabla_{\sigma}A^{\delta}
\right]\,.
\label{SGBf}
\ee
This belongs to a subclass of beyond GP theories 
proposed in Ref.~\cite{Heisenberg:2016eld} and 
thus it is free from the Ostrogradski ghost.

In theories where the function $f(X)$ is constant, i.e., 
\be
f(X)=\alpha={\rm constant}\,,
\label{fc}
\ee
we have 
\be
G_5(X)=4\alpha \ln |X|\,,\qquad 
f_5(X)=0\,.
\label{G5v}
\ee
Note that we dropped a constant term $8\alpha$ in $G_5(X)$, 
as it does not contribute to the field equations of motion.
Since the last term on the right hand side of Eq.~(\ref{SGBf}) vanishes, the action ${\cal S}_{\rm vGB}$ belongs to a subclass of 
GP theories \cite{Heisenberg:2014rta,Tasinato:2014eka,BeltranJimenez:2016rff}. 
We recall that, from Eq.~(\ref{Ggauss}), the linear scalar-GB coupling $\mu(\phi) \mG$ 
with $\mu(\phi)=-\alpha \phi$ corresponds to $G_{5s}(X_s)=4\alpha \ln |X_s|$ 
with $G_{2s}(X_s)=G_{3s}(X_s)=G_{4s}(X_s)=0$.
The coupling (\ref{G5v}) is the vector-tensor analogue to the linear 
scalar-GB coupling in Horndeski theories.

In the following, we will focus on GP theories given by 
the functions (\ref{G5v}).
Varying the action (\ref{SGBf}) with respect to $A_{\mu}$, 
it follows that 
\begin{align}
\frac{\delta {\cal S}_{\rm vGB}}{\delta A_{\mu}}
=\alpha \left( \mathcal{J}^{\mu} + \mathcal{J}_F^{\mu} \right)\,,
\end{align}
where $\mathcal{J}^{\mu}$ is defined by Eq.~(\ref{Jmu}), and 
\begin{align}
\mathcal{J}_F^{\mu} \equiv \delta^{\mu\nu\rho\sigma}_{\alpha\beta\gamma\delta}
A_{\nu}F^{\alpha\beta}\left( \frac{1}{X^2}
\nabla^{\gamma}A_{\rho} \nabla^{\delta}A_{\sigma} 
-\frac{1}{2X}R^{\gamma\delta}{}_{\rho\sigma}\right)\,,
\end{align}
with $F^{\alpha \beta} \equiv \nabla^{\alpha}A^{\beta}-\nabla^{\beta}A^{\alpha}$.
The sum of $\mathcal{J}^{\mu}$ and $\mathcal{J}_F^{\mu}$ can be expressed in a compact form
\be
\mathcal{J}^{\mu} + \mathcal{J}_F^{\mu}
=\delta^{\mu\nu\rho\sigma}_{\alpha\beta\gamma\delta}\left[ \frac{A^{\alpha} 
\nabla^{\beta}A_{\nu}}{X} \left( R^{\gamma\delta}{}_{\rho\sigma}
-\frac{2}{3X}\nabla^{\gamma}A_{\rho}\nabla^{\delta}A_{\sigma} \right) 
\right]\,.
\ee

Let us consider the case in which the Maxwell term 
$F \equiv -(1/4)F^{\alpha \beta}F_{\alpha \beta}$ 
and the mass term $X$ are present in addition to 
the action (\ref{SGBf}) with $f(X)=\alpha$.
The action in such a subclass of GP theories is given by 
\be
{\cal S}=\int {\rm d}^4 x \sqrt{-g} \left[ \frac{1}{g_{\rm v}^2}F+\eta X
+G_5(X) G_{\mu\nu}\nabla^{\mu}A^{\nu} -\frac{1}{6}G_{5,X}(X)
\delta^{\mu\nu\rho}_{\alpha\beta\gamma}\nabla_{\mu}A^{\alpha
}\nabla_{\nu}A^{\beta} \nabla_{\rho}A^{\gamma} 
\right]\,,
\label{action0}
\ee
where $g_{\rm v}$ and $\eta$ are constants
and $G_5(X)=4\alpha \ln |X|$. 
Varying this action with respect to $A_{\mu}$, it follows that 
\be
\frac{1}{g_{\rm v}^2}\nabla_{\nu}F^{\mu\nu}+\eta A^{\mu}
=\alpha (\mathcal{J}^{\mu} + \mathcal{J}_F^{\mu})\,. 
\label{eom_vGB}
\ee
While the GB term does not explicitly show up in Eq.~(\ref{eom_vGB}), 
it appears by taking the divergence of Eq.~(\ref{eom_vGB}) as
\be
\eta \nabla_{\mu} A^{\mu} = \alpha \mathcal{G} 
+\alpha \nabla_{\mu} \mathcal{J}_F^{\mu}\,, 
\label{constraint_vGB}
\ee
where we used the relation $\nabla_{\mu}\nabla_{\nu}F^{\mu\nu}=0$ 
and Eq.~(\ref{Jmu2}). 

Taking the decoupling limit $g_{\rm v} \to 0$ with 
the replacement $A^{\mu} \to g_{\rm v} A^{\mu} + \nabla^{\mu}\phi$, 
we have $\mathcal{J}_F^{\mu} \to 0$ and hence 
Eqs.~\eqref{eom_vGB} and \eqref{constraint_vGB} reduce to 
the Maxwell equation, $\nabla_{\nu}F^{\mu\nu}=0$, and 
the equation of motion for the scalar field, 
$\eta \nabla_{\mu} \nabla^{\mu} \phi=\alpha \mG$, respectively. 
This latter scalar field equation also follows by varying the Lagrangian 
$L=\eta X_s-\alpha \phi \mG$ with respect to $\phi$, so
the linearly coupled scalar-GB theory is recovered 
by taking the above decoupling limit.
In shift-symmetric Horndeski theories, using the expression of $\mG$ in Eq.~(\ref{mG}) 
shows that the scalar field equation can be expressed in the form 
$\nabla_{\mu}j^{\mu}=0$, where $j^{\mu}$ is a conserved current. 
When this equation is solved for $j^{\mu}$, there is an integration constant 
corresponding to boundary/initial conditions of the system. 

In vector-tensor theories the equation of motion for the vector field $A^{\mu}$ 
is given by Eq.~(\ref{eom_vGB}), which does not contain an integration constant.  
Although Eq.~(\ref{constraint_vGB}) corresponds to the differential version 
of Eq.~(\ref{eom_vGB}), one cannot choose an arbitrary integration constant 
when integrating Eq.~(\ref{constraint_vGB}).
This property is different from that in shift-symmetric Horndeski
theories discussed above. 
If we apply GP theories to the isotropic and homogeneous cosmological 
background, the temporal vector component $A_0$ is always related to the Hubble 
expansion rate $H$ \cite{DeFelice:2016yws,DeFelice:2016uil,deFelice:2017paw}. 
This is known as a tracker solution, in which case we do not have 
a freedom of changing initial conditions of the vector field. 
In scalar-tensor theories, on the other hand, it is possible to choose initial conditions 
away from the tracker because of the existence of the integration 
constant said above \cite{DeFelice:2010pv,DeFelice:2011bh}. 
As a result, one can distinguish between 
shift-symmetric Horndeski theories and GP theories from 
the background cosmological dynamics.

If we apply linear scalar-GB theory to the static and spherically symmetric 
background in vacuum, there are hairy BH solutions satisfying the boundary condition 
$X_s=0$ on the horizon \cite{Sotiriou:2013qea,Sotiriou:2014pfa}. 
This boundary condition fixes the integration constant 
mentioned above \cite{Minamitsuji:2022mlv}. 
In vector-GB theory, the similar boundary condition, like $X=0$, cannot be 
necessarily imposed because of the absence of an arbitrary constant in Eq.~(\ref{eom_vGB}). 
This implies that the BH solution in vector-GB theory should be different from 
that in linear scalar-GB theory. 
In Sec.~\ref{BHsec}, we will investigate the property of hairy 
BH solutions in vector-GB theory.

\section{Black hole solutions in vector-GB theory}
\label{BHsec}

We study BH solutions by incorporating the 
Einstein-Hilbert term in the action (\ref{action0}). 
The corresponding action belongs to a subclass of GP 
theories given by 
\begin{align}
{\cal S}&=
\int {\rm d}^4 x \sqrt{-g} \left[ \frac{\Mpl^2}{2}R
+F+\eta X + \alpha A_{\mu} {\cal J}^{\mu} \right]
\nonumber \\
&=
\int {\rm d}^4 x \sqrt{-g} \left[ \frac{\Mpl^2}{2}R
+F+\eta X
+(4\alpha \ln |X|) G_{\mu\nu}\nabla^{\mu}A^{\nu}
-\frac{2\alpha}{3X}
\delta^{\mu\nu\rho}_{\alpha\beta\gamma}\nabla_{\mu}A^{\alpha
}\nabla_{\nu}A^{\beta} \nabla_{\rho}A^{\gamma} 
\right]\,,
\label{action}
\end{align}
where $\Mpl$ is the reduced Planck mass 
and we set $g_{\rm v}=1$.
Let us consider the static and spherically symmetric 
background given by the line element 
\be
{\rm d}s^{2} =-f(r) {\rm d}t^{2} +h^{-1}(r) {\rm d}r^{2} + 
r^{2} \left( {\rm d}\theta^{2}
+\sin^{2}\theta\, {\rm d} \varphi^{2} \right)\,,
\label{spmetric}
\ee
where $t$, $r$ and $(\theta,\varphi)$ represent the time, radial,
and angular coordinates, respectively, 
and $f$ and $h$ are functions of $r$. 
We will focus on the case of positive mass squared $\eta>0$ 
as in standard massive Proca theory.
For the vector field, we consider the following configuration 
\be
A_{\mu}=\left[ A_0 (r), A_1(r), 0, 0 \right]\,,
\ee
where $A_0$ and $A_1$ are functions of $r$. 
On the background (\ref{spmetric}), we have 
\be
X=\frac{A_0^2}{2f}-\frac{hA_1^2}{2}\,,\qquad 
F=\frac{hA_0'^2}{2f}\,,
\ee
where a prime represents the derivative with respect to $r$. 
Note that $F=-F_{\mu\nu}F^{\mu\nu}/4 \neq 0$ implies a nonvanishing 
temporal vector component; that is, $A_{\mu}$ cannot be expressed by a scalar gradient. Hence, the presence of a nontrivial temporal component $A_0(r)$ is essential to differentiate solutions in vector-GB theory from those in scalar-GB theory.

Varying the action (\ref{action}) with respect to $f$ and $h$, 
respectively, we obtain
\ba
& &
\Mpl^2 f \left( r h'+h-1 \right)+\frac{r^2 h A_0'^2}{2}
+\frac{\eta r^2}{2} \left( A_0^2+fh A_1^2 \right)
+\frac{4\alpha f}{(A_0^2-fh A_1^2)^2} 
[2A_0^3A_0'A_1h(h - 1) + 2A_0 A_0'A_1^3 
f h^2 (h + 1) 
\nonumber \\
& &
+A_0^4 \{ 2A_1' h(h - 1) + A_1 h'(3h - 1)\} 
+A_1^4 f^2 h^2 \{ 2A_1' h (h - 1) + A_1 h'(3h - 1)\} \nonumber \\
& &
+2A_0^2 A_1^2 f h \{ A_1 h'(1-3h)
-2A_1' h(h - 1)\}]=0\,,\label{heq0} \\
& &
\Mpl^2 \left[ r hf'+f(h-1) \right]+\frac{r^2 h A_0'^2}{2}
-\frac{\eta r^2}{2} \left( A_0^2+fh A_1^2 \right)
+\frac{4\alpha h A_1}{(A_0^2-fh A_1^2)^2} 
[ 2A_0' A_0^3 f(1-3h)+2A_0^2 A_1^2 f' f h (1-3h)\nonumber \\
& &
+2A_0' A_0 A_1^2 f^2 h (h-1)
+A_0^4 f' (3h-1)+A_1^4 f' f^2 h^2 (3h-1)]=0\,.
\label{feq0}
\ea
Variations of the action (\ref{action}) with respect to $A_0$ and 
$A_1$ lead, respectively, to 
\ba
\hspace{-1cm}
& &
A_0''+\left( \frac{2}{r}-\frac{f'}{2f}+\frac{h'}{2h} \right)A_0'
-\frac{\eta}{h}A_0-\frac{4\alpha A_0}{r^2 h (A_0^2-f h A_1^2)^2}
[ A_0^2 \{ A_1f' h(h-1)+2A_1' fh (h-1)+A_1 f h' (3h-1)\} \nonumber \\
\hspace{-1cm}
& &+A_1^2 fh \{ A_1 f' h (h+1)+2A_1' fh (h+1)+A_1 fh'(1-h)\}]=0\,,
\label{A0eq} \\
\hspace{-1cm}
& &
\eta A_1 
-\frac{4\alpha}{r^2 f  (A_0^2-f h A_1^2)^2} \left[ (h-1) 
(f' f^2 h^2 A_1^4-2f' fh A_0^2 A_1^2+f'A_0^4
-2fA_0' A_0^3)-2f^2 h(h+1) A_1^2 A_0' A_0 
\right]=0\,.
\label{A1eq}
\ea

In the absence of the vector-GB coupling ($\alpha=0$) with $\eta \neq 0$, 
we have $A_1(r)=0$ from Eq.~(\ref{A1eq}). 
For the asymptotically flat boundary conditions where $f$ and $h$ 
approach $1$ at spatial infinity, the large-distance 
solution to Eq.~(\ref{A0eq}) is given by 
$A_0=C_1 e^{-\sqrt{\eta} r}/r+C_2 e^{\sqrt{\eta} r}/r$. 
To avoid the divergence of $A_0$ at spatial infinity, 
we have to choose $C_2=0$ and hence $A_0=C_1 e^{-\sqrt{\eta} r}/r$ at 
large distances. 
{}From Eqs.~(\ref{heq0}) and (\ref{feq0}), 
we obtain the relation $(f/h)'=\eta A_0^2 r/(\Mpl^2 h^2)$.
Since $f/h$ should be a finite constant on the BH horizon, 
we need to require that $A_0=0$.
Indeed, the solution consistent with all the background equations 
and boundary conditions is $A_0(r)=0$ at any radius. 
In this case, we end up with the Schwarzschild solution 
with the metric components $f=h=1-r_h/r$, where $r_h$ 
is the horizon radius.

For $\alpha \neq 0$, Eq.~(\ref{A1eq}) shows that 
it is possible to realize the solution with $A_1 \neq 0$.
Moreover, the nonvanishing radial component $A_1$ affects the 
temporal component $A_0$ through the $\alpha$-dependent terms 
in Eq.~(\ref{A0eq}). For simplicity, 
we shall seek solutions with $A_{\mu}\neq 0$ 
for a small coupling constant $\alpha$ and leave general analysis for future work. When $\alpha=0$ we only have the trivial solution $A_{\mu}=0$, so the solutions for a small $\alpha$ may be scaled as 
$A_{\mu}=\mathcal{O}(\alpha)$.
Let us express the leading-order solutions to 
$A_0$ and $A_1$ in the forms
\be
A_0(r)=\alpha \tilde{A}_0(r)\,,\qquad 
A_1(r)=\alpha \tilde{A}_1(r)\,,
\ee
where $\tilde{A}_0$ and $\tilde{A}_1$ are functions of $r$.
Then, from Eqs.~(\ref{A0eq}) and (\ref{A1eq}), we obtain
\ba
\hspace{-1cm}
& &
\tilde{A}_0''+\left( \frac{2}{r}-\frac{f'}{2f}+\frac{h'}{2h} \right)\tilde{A}_0'
-\frac{\eta}{h}\tilde{A}_0-\frac{4 \tilde{A}_0}
{r^2 h (\tilde{A}_0^2-f h \tilde{A}_1^2)^2}
[ \tilde{A}_0^2 \{ \tilde{A}_1f' h(h-1)+2\tilde{A}_1' fh (h-1)
+\tilde{A}_1 f h' (3h-1)\} \nonumber \\
\hspace{-1cm}
& &+\tilde{A}_1^2 fh \{ \tilde{A}_1 f' h (h+1)
+2\tilde{A}_1' fh (h+1)+\tilde{A}_1 fh'(1-h)\}]=0\,,
\label{A0eqa} \\
\hspace{-1cm}
& &
\eta \tilde{A}_1 
-\frac{4}{r^2 f  (\tilde{A}_0^2-f h \tilde{A}_1^2)^2} \left[ (h-1) 
(f' f^2 h^2 \tilde{A}_1^4-2f' fh \tilde{A}_0^2 \tilde{A}_1^2
+f'\tilde{A}_0^4
-2f\tilde{A}_0' \tilde{A}_0^3)-2f^2 h(h+1) 
\tilde{A}_1^2 \tilde{A}_0' \tilde{A}_0 
\right]=0\,.
\label{A1eqa}
\ea
As we observe in Eqs.~(\ref{heq0}) and (\ref{feq0}), 
the vector field contributions to metric components $f$ and $h$ 
arise at second order in $\alpha$.
Then, up to first order in $\alpha$, we can exploit the 
Schwarzschild metric components:
\be
f=h=1-\frac{r_h}{r}\,.
\label{fh}
\ee
We substitute Eq.~(\ref{fh}) and its derivatives into 
Eqs.~(\ref{A0eqa}) and (\ref{A1eqa}).

Note that we only need to impose two boundary conditions to solve 
Eqs.~\eqref{A0eqa} and \eqref{A1eqa}. 
Although Eq.~\eqref{A0eqa} contains a second-order derivative of $\tilde{A}_0$, one can express $\tilde{A}_0''$ with respect to first-order derivatives of $\tilde{A}_0$ and $\tilde{A}_1$ by differentiating Eq.~\eqref{A1eqa}. 
Then, one obtains a set of first-order differential equations of $\tilde{A}_0$ and $\tilde{A}_1$.

\subsection{Boundary conditions}

Around the horizon, we expand the temporal vector component 
in the form 
\be
\tilde{A}_0=\sum_{i=0}a_i \left( \frac{r-r_h}{r_h} \right)^i
=a_0+a_1 \frac{r-r_h}{r_h}+a_2\frac{(r-r_h)^2}{r_h^2}\cdots\,, 
\label{A0expan}
\ee
where $a_i$'s are constants. 
If $\tilde{A}_0$ decreases around $r=r_h$, $a_1$ is negative.
We are interested in regular vector field solutions where 
both $X$ and $F$ are finite on the horizon. 
The leading-order contribution to $F$ at $r=r_h$ is 
$a_1^2/(2r_h^2)$. 
To keep $X$ finite on the horizon, we require that 
the leading-order radial vector component diverges as 
$\tilde{A}_1=\tilde{A}_0/\sqrt{fh}=a_0 r/(r-r_h)$ at $r=r_h$. 
In this case, we can expand $\tilde{A}_1$ in the form 
\be
\tilde{A}_1=a_0 \frac{r}{r-r_h}
+\sum_{i=0} b_i \left( \frac{r-r_h}{r_h} \right)^i
=a_0 \frac{r}{r-r_h}+b_0+b_1 \frac{r-r_h}{r_h}\cdots\,, 
\label{A1expan}
\ee
where $b_i$'s are constants. 
Then, we have $X=a_0(a_1-b_0)\alpha^2+{\cal O}(r-r_h)$ 
in the vicinity of $r=r_h$. 
Substituting Eqs.~(\ref{A0expan})-(\ref{A1expan}) into 
Eqs.~(\ref{A0eqa})-(\ref{A1eqa}), we find that $b_0$ and $b_1$ 
are related to $a_0$, $a_1$, and $a_2$ according to 
\ba
b_0 &=&
\frac{2r_m^2 a_1+ r_h^3 a_0 a_1
\pm 2\sqrt{r_m^2 a_1[(a_1-4a_0)r_m^2-r_h^3 a_0^2 ]}}{4r_m^2+r_h^3 a_0}\,,
\label{b0re}\\
b_1 &=&
[ 2(a_1 - b_0)^3 a_1 r_m^2  + 2a_0^2(6a_1^2 - 2a_1 b_0 + 4a_2 b_0)  r_m^2 
+ 4 a_0 (a_1 - b_0) \{ 
a_1^2 + a_1(a_2 - 7b_0) - 2(a_2 - 2b_0)b_0 \} r_m^2 
\nonumber \\
&& -a_0(a_1 - b_0)^3 (a_0 + b_0)r_h^3]
/[4 a_0 a_1 (2 a_0 - a_1 + b_0) r_m^2]\,,
\label{b1re}
\label{b0}
\ea
where we set 
\be
\eta \equiv \frac{1}{r_m^2}\,.
\ee

We would like to derive asymptotic solutions of
$\tilde{A}_0$ and $\tilde{A}_1$ approaching 0 
at spatial infinity. 
Note that the GB corrections to \eqref{A0eqa} and \eqref{A1eqa} are ``zeroth'' order in $A_{\mu}$, that is, the power of $A_{\mu}$ in the denominator and that in the numerator are the same. 
Therefore, the limit $A_{\mu} \to 0$ needs to be carefully analyzed. 
Since the equation of motion for $\tilde{A}_0$ contains a mass term 
$\tilde{A}_0/r_m^2$, we search for solutions where $\tilde{A}_0$ 
decreases as $e^{-r/r_m}/r$ or faster while $\tilde{A}_1$ 
decreases slower than $\tilde{A}_0$. 
Ignoring the $\tilde{A}_0$-dependent 
contributions to Eq.~(\ref{A1eqa}), it follows that 
\be
\frac{\tilde{A}_1}{r_m^2}
-\frac{4(h-1)f'}{r^2f}
= 0\,.
\ee
Then, the radial vector component should have the 
asymptotic behavior
\be
\tilde{A}_1=-\frac{4r_h^2 r_m^2}{r^5}\,,
\label{A1as}
\ee
whose amplitude decreases in proportion to $r^{-5}$.
For the last terms of Eq.~(\ref{A0eqa}) containing the 
squared bracket, we neglect 
the $\tilde{A}_0$-dependent contributions and 
substitute the solution (\ref{A1as}) into Eq.~(\ref{A0eqa}).
Then, at large distances, we have
\be
\tilde{A}_0''+\frac{2}{r} 
\tilde{A}_0'-\frac{\tilde{A}_0}{r_m^2}
+\frac{8r}{r_m^2 r_h}\tilde{A}_0 
-\frac{20r^2}{r_h^2 r_m^2}\tilde{A}_0=0\,.
\ee
Ignoring the third and fourth terms relative to the fifth one 
in the regime $r\gg r_h$, we obtain the following asymptotic 
solution 
\be
\tilde{A}_0=\frac{C_1}{\sqrt{r}} I_{1/4}\left(\frac{\sqrt{5}r^2}{r_h r_m} \right)+
\frac{C_2}{\sqrt{r}} K_{1/4}\left(\frac{\sqrt{5}r^2}{r_h r_m} \right)\,,
\label{A0geasy}
\ee
where $I_{1/4}$ and $K_{1/4}$ are the Bessel functions of 
first and second kinds, respectively, and $C_1$, 
$C_2$ are integration constants. 
The boundary condition avoiding the divergence of 
$\tilde{A}_0$ corresponds to $C_1=0$ and hence
\be
\tilde{A}_0=
\frac{C_2}{\sqrt{r}} K_{1/4}\left(\frac{\sqrt{5}r^2}{r_h r_m} \right)\,,
\label{A0as}
\ee
which decreases as 
$\tilde{A}_0 \propto 
r^{-3/2} 
e^{-\sqrt{5}r^2/(r_h r_m)}$. Note that this solution decreases even 
faster than $e^{-r/r_m}/r$, 
but the discussion for deriving Eqs.~(\ref{A1as}) and 
(\ref{A0as}) does not lose 
its validity. Therefore, we have obtained a consistent asymptotic solution \eqref{A1as} and \eqref{A0as} which contains one parameter undetermined by the asymptotic boundary condition $A_{\mu} \to 0$.

\subsection{Numerical solutions}

In this section, we will numerically study the existence of 
hairy BH solutions in theories given by the action (\ref{action}).
For this purpose, we introduce a new variable $\tilde{B}_1$ 
defined by 
\begin{align}
\tilde{B}_1 \equiv \tilde{A}_1-\frac{\tilde{A}_0}{f}  
\,, \label{tB1}
\end{align}
with $f=1-r_h/r$.
Around the horizon, using the expanded solutions (\ref{A0expan}) 
and (\ref{A1expan}) leads to 
\be
\tilde{B}_1=b_0-a_1+(b_1-a_1-a_2) \frac{r-r_h}{r_h}
+{\cal O}\left( \frac{(r-r_h)^2}{r_h^2} \right)\,.
\label{tB1d}
\ee
Unlike $\tilde{A}_1$, the new variable has a finite 
value $\tilde{B}_1(r_h)=b_0-a_1$ on the horizon.
We also have
\begin{align}
X= -\frac{\alpha^2}{2} \tilde{B}_1(2\tilde{A}_0+f \tilde{B}_1)\,,
\label{Xge}
\end{align}
which manifests the regularity of $X$ for finite 
values of $\tilde{A}_0$ and $\tilde{B}_1$.

Taking the $r$ derivative of Eq.~(\ref{A1eqa}) and using Eq.~(\ref{A0eqa}) 
to eliminate $\tilde{A}_0''$, we obtain the first-order differential 
equation containing $\tilde{A}_1'$. 
We also note that Eq.~(\ref{A1eqa}) can be regarded as the first-order 
coupled differential equation for $\tilde{A}_0$.
Then, the background equations can be expressed in the forms
\begin{align}
\tilde{A}_0'=F_0[\tilde{A}_0, \tilde{B}_1]\,, 
\label{A0deq} \\
\tilde{B}_1'=F_1[\tilde{A}_0, \tilde{B}_1]\,,
\label{B1deq} 
\end{align}
where the functions $F_0$ and $F_1$ depend on $\tilde{A}_0$ 
and $\tilde{B}_1$. The right hand sides of Eqs.~(\ref{A0deq}) 
and (\ref{B1deq}) are regular on the horizon. 

At large distances ($r \gg r_h$), $\tilde{A}_0$ decreases much faster than 
$\tilde{A}_1$ and hence $X \simeq -\alpha^2 \tilde{B}_1^2/2<0$, where 
$\tilde{B}_1 \simeq \tilde{A}_1=-4r_h^2 r_m^2/r^5$. 
Since the background Eqs.~(\ref{A0eqa}) and (\ref{A1eqa}) contain 
terms proportional to $X$ in their denominators, the regularity of 
these equations means that $X$ should not change its sign 
throughout the horizon exterior, i.e., $X<0$ for $r>r_h$. 
In the vicinity of $r=r_h$, we have 
\be
X=\alpha^2 \frac{2a_0 [a_1 r_m^2 
\mp \sqrt{a_1^2 r_m^4-a_0 a_1 r_m^2(4r_m^2+a_0 r_h^3)}]}
{4r_m^2+a_0 r_h^3}+{\cal O} \left( \frac{r-r_h}{r_h} \right)\,,
\label{Xrh}
\ee
where the minus and plus signs correspond to the 
plus and minus signs of $b_0$ in Eq.~(\ref{b0re}), respectively.
If we choose the minus branch of Eq.~(\ref{Xrh}), we have $X<0$ for 
$a_0>0$ and $a_1<0$. 
For the plus branch of Eq.~(\ref{Xrh}), it is possible to realize 
$X<0$ for $a_0<0$ and $a_1>0$, so long as the condition
$4r_m^2+a_0 r_h^3>0$ is satisfied. 

Around the horizon, we have $\tilde{A}_1>0$ and $\tilde{A}_1'<0$ 
for $a_0>0$ from Eq.~(\ref{A1expan}), whereas, for $a_0<0$, 
$\tilde{A}_1<0$ and $\tilde{A}_1'>0$. 
At spatial infinity, Eq.~(\ref{A1as}) gives 
$\tilde{A}_1<0$ and $\tilde{A}_1'>0$. 
Then, the branch smoothly matching the solutions 
in two asymptotic regimes ($r \simeq r_h$ and $r \gg r_h$) 
without the sign changes of $\tilde{A}_1$ and $\tilde{A}_1'$ 
corresponds to the minus sign of $b_0$.
We will search for numerical solutions in this latter regime, 
i.e., $a_0<0$, $a_1>0$, and $4r_m^2+a_0 r_h^3>0$. 
In this case, we have $b_0<0$ from Eq.~(\ref{b0re}).
Around $r=r_h$, the variable $\tilde{B}_1$  
behaves as Eq.~(\ref{tB1d}),  
whose leading-order term $b_0-a_1$ is negative. 
The first-order coupled differential Eqs.~(\ref{A0deq}) and (\ref{B1deq}) 
can be integrated outward for two given boundary conditions of 
$\tilde{A}_0(r_h)=a_0$ and $\tilde{B}_1(r_h)=b_0-a_1$. 

Since $b_0$ is expressed by using $a_0$ and $a_1$ as 
Eq.~(\ref{b0re}), choosing two boundary conditions on the horizon 
amounts to fixing the two constants $a_0$ and $a_1$ in the expansion of 
$\tilde{A}_0$ given by Eq.~(\ref{A0expan}), with 
$\tilde{A}_1=a_0 r/(r-r_h)+b_0$.
While we need two boundary conditions on the horizon, 
there is only one undetermined integration constant $C_2$ 
at spatial infinity. Two boundary conditions at $r=r_h$ may not be 
uniquely fixed even if we impose the regularity on the horizon 
and $A_{\mu} \to 0$ at spatial infinity. 
This suggests the existence of a family of solutions, 
which we will address in the following. 

\vspace{0.5cm}
\begin{figure}[h]
\begin{center}
\includegraphics[width=0.45\linewidth]{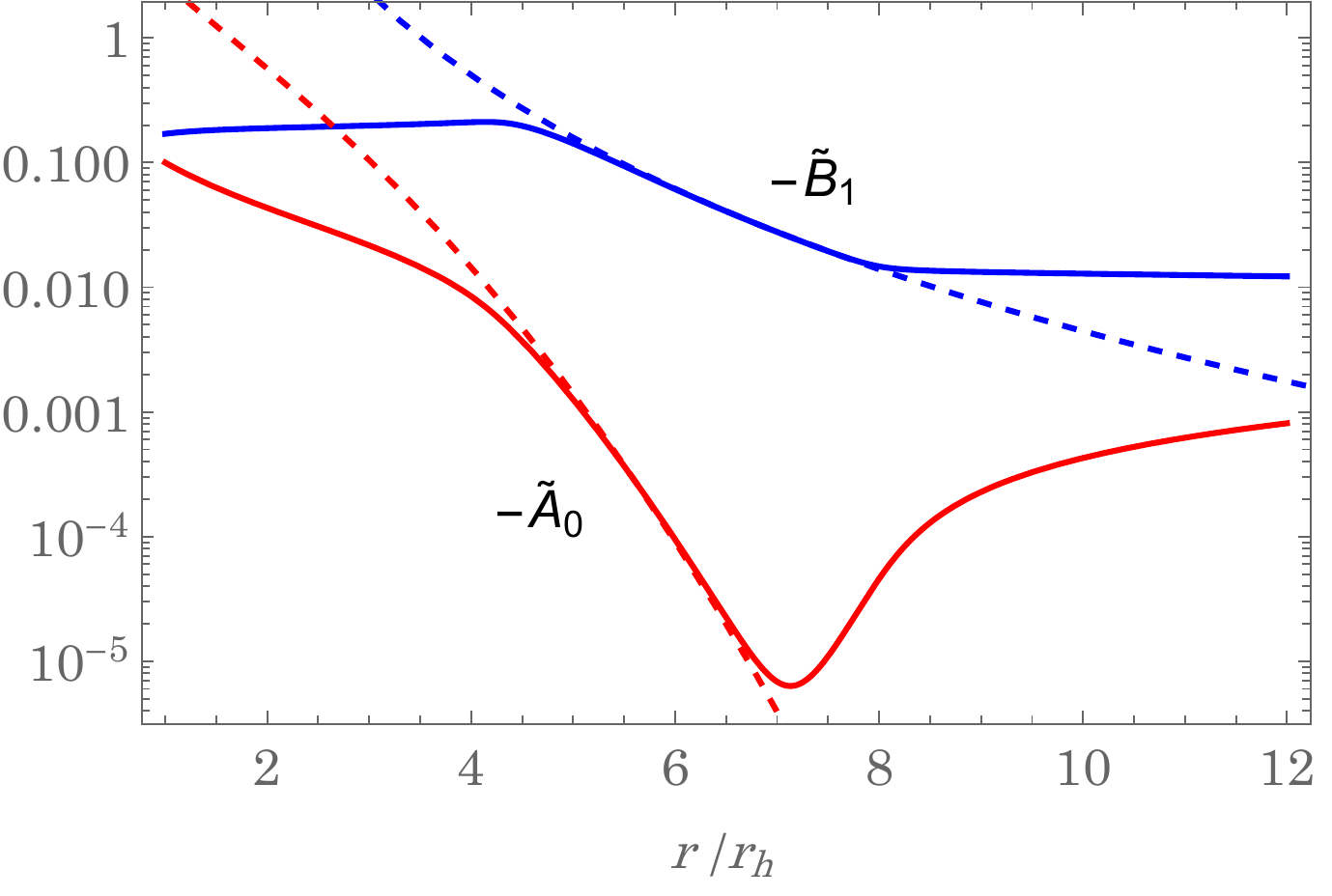}
\hspace{0.2cm}
\includegraphics[width=0.45\linewidth]{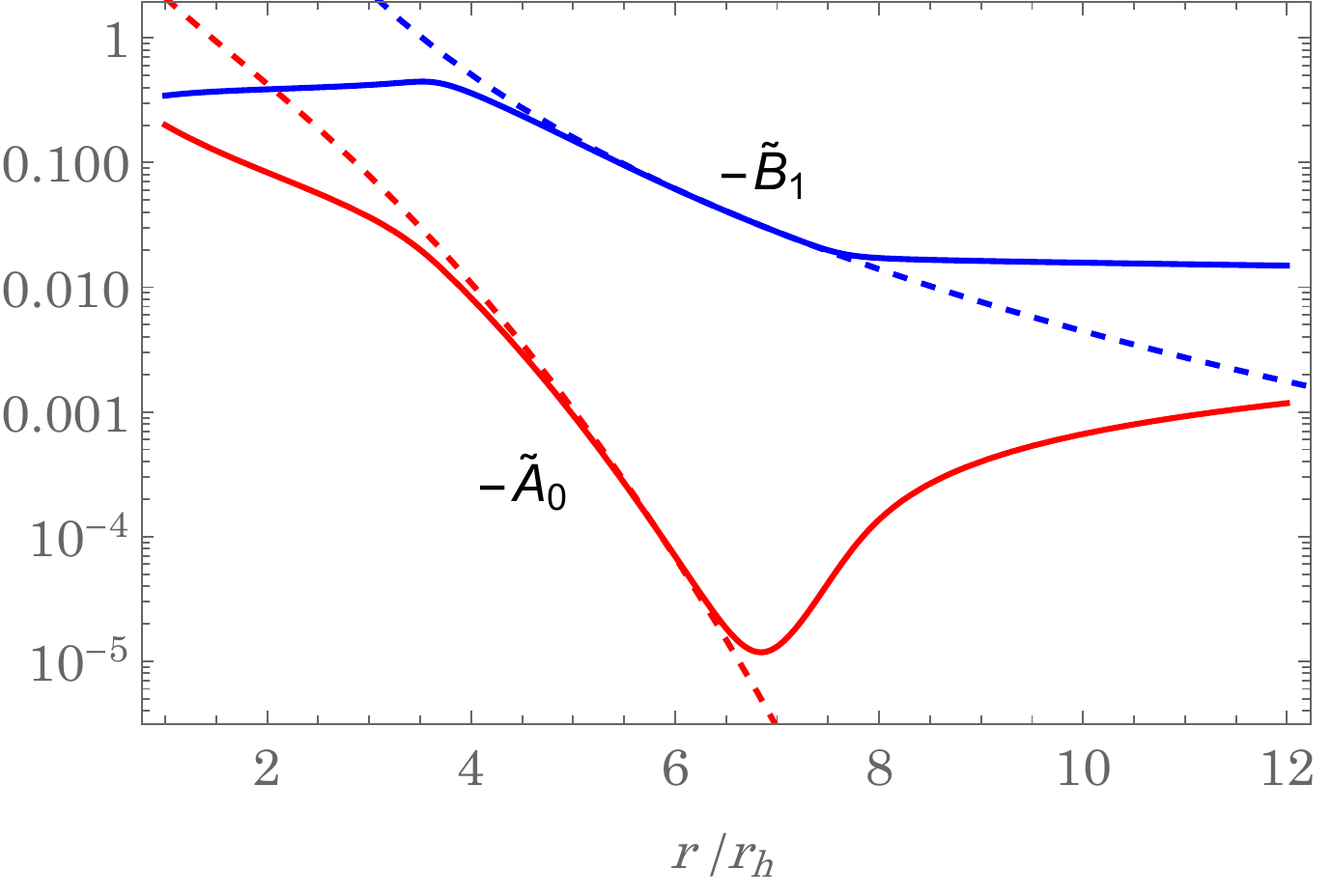}
\end{center}
\caption{\label{fig1} 
Numerically derived solutions to $-\tilde{A}_0$ (red solid) 
and $-\tilde{B}_1$ (blue solid) versus $r/r_h$ for 
$r_m/r_h=10$ with the boundary conditions
(i) $\tilde{A}_0(r_h)=-0.1/r_h$, 
$\tilde{B}_1(r_h)=-0.17026504/r_h$ (left), and 
(ii) $\tilde{A}_0(r_h)=-0.2/r_h$, 
$\tilde{B}_1(r_h)=-0.34443001/r_h$ (right). 
Both $-\tilde{A}_0$ and  $-\tilde{B}_1$ are normalized 
by $1/r_h$.
The dashed red and blue curves represent the large-distance 
analytic solutions to $-\tilde{A}_0$ and $-\tilde{B}_1$ 
determined by Eqs.~(\ref{A0as}) and (\ref{A1as}), 
respectively.
}
\end{figure}

Numerically, it is difficult (if possible) to find solutions satisfying $A_{\mu} \to 0$ at spatial infinity because the asymptotic solution generically has a nonvanishing growing mode. Instead, we find solutions that smoothly connect to their large-distance analytic solutions at a finite distance by using the shooting method. 
For this purpose, we fix $r_m=10r_h$ in our numerical simulations.
In this case, unless the boundary conditions on the horizon are carefully chosen, the growing-mode solution of $\tilde{A}_0$, which corresponds to 
the first term on the right hand side of Eq.~(\ref{A0geasy}), 
should manifest itself for the distance 
$r \gtrsim (10/\sqrt{5})^{1/2}=1.4 r_h$. Under appropriate boundary conditions, on the other hand, the solutions can match the large-distance analytic solutions at least at certain distances, which we shall regard as our numerical solutions.

In the left panel of Fig.~\ref{fig1}, we plot the numerically derived 
values of $-\tilde{A}_0$ and $-\tilde{B}_1$ versus $r/r_h$ 
as solid lines for $r_m/r_h=10$ with the boundary conditions 
$\tilde{A}_0(r_h)=-0.1/r_h$ and $\tilde{B}_1(r_h)=-0.17026504/r_h$. 
The analytic solutions of $-\tilde{A}_0$ and $-\tilde{B}_1$ 
at large distances are also shown as dashed lines, which are 
obtained from Eqs.~(\ref{A0as}) and (\ref{A1as}) with the choice $C_2=-1.6$. 
We observe that both $-\tilde{A}_0$ and $-\tilde{B}_1$ approach 
these large-distance solutions around the distance 
$r \gtrsim 4.5 r_h$. We also find that $\tilde{A}_0$ starts to 
deviate from the decaying mode (\ref{A0as}) for $r \gtrsim 6.5 r_h$, 
which is followed by the departure of $\tilde{B}_1$ from 
the large-distance solution $-4r_h^2 r_m^2/r^5$. 
This behavior is attributed to the presence of a growing mode  
$(C_1/\sqrt{r})I_{1/4}(\sqrt{5}r^2/(r_h r_m))$ in $\tilde{A}_0$. 
Since such a rapidly growing mode is very sensitive to the 
accumulation of tiny numerical errors, 
it is challenging to find the exact boundary 
conditions at $r=r_h$ realizing $C_1=0$ at spatial infinity. 

However, as we have mentioned, the fact that there are regions of the distance in which 
$\tilde{A}_0$ and $\tilde{B}_1$ can be well approximated by 
their large-distance analytic expressions means that solutions with
the proper asymptotic behavior should exist for the boundary conditions 
close to $\tilde{A}_0(r_h)=-0.1/r_h$ and $\tilde{B}_1(r_h)=-0.17026504/r_h$.  
If we fix $\tilde{A}_0(r_h)=-0.1/r_h$ and vary $\tilde{B}_1(r_h)$, 
we have not found other ranges of $\tilde{B}_1(r_h)$
in which $-\tilde{A}_0$ and $-\tilde{B}_1$ temporally 
approach their large-distance solutions like the left panel of Fig.~\ref{fig1}. 
If we consider other boundary conditions of $\tilde{A}_0(r_h)$ 
around $-0.1/r_h$, there are solutions which can be well approximated by the 
large-distance solutions for some ranges of $r$. 
The right panel of Fig.~\ref{fig1} corresponds to such an example, 
in which case $\tilde{A}_0(r_h)=-0.2/r_h$ and 
$\tilde{B}_1(r_h)=-0.34443001/r_h$. 
We also found similar cases for $\tilde{A}_0(r_h)$ larger than 
$-0.1/r_h$, say $\tilde{A}_0(r_h)=-0.09/r_h$, 
by choosing the values of $\tilde{B}_1(r_h)$ properly. 
These facts show that there are appropriate solutions of 
$\tilde{A}_0$ and $\tilde{B}_1$ connecting two 
asymptotic regimes ($r \simeq r_h$ and $r \gg r_h$) for 
some ranges of $\tilde{A}_0(r_h)$ around $-0.1/r_h$. 
If $\tilde{A}_0(r_h)$ is far away from the order $-0.1/r_h$, 
it is typically difficult to find the parameter spaces of 
$\tilde{A}_0(r_h)$ and $\tilde{B}_1(r_h)$ in which 
both $\tilde{A}_0$ and $\tilde{B}_1$ approach 
their asymptotic solutions. 

We thus showed that, for $r_m=10r_h$, there are some ranges of 
$\tilde{A}_0(r_h)$ and $\tilde{B}_1(r_h)$ in which the solutions 
in two asymptotic regimes can be smoothly connected. 
This means that the solutions are not uniquely fixed even for 
a fixed vector mass term. 
From Eq.~(\ref{A1expan}), the radial vector component $\tilde{A}_1$ 
diverges at $r=r_h$. In massless scalar-tensor theories with 
the linear scalar-GB coupling $-\alpha \phi{\cal G}$, 
the field derivative $\phi'$ for linearly stable hairy 
BH solutions is finite on the horizon and hence 
$X_s=-(1/2)h \phi'^2$ vanishes there \cite{Minamitsuji:2022mlv}. 
In the latter case, the BH solution with $X_s(r_h)=0$  
is uniquely fixed by performing the expansions of scalar field 
and metrics with respect to the small coupling 
$\alpha$ \cite{Sotiriou:2013qea,Sotiriou:2014pfa,Minamitsuji:2022mlv}.
It is also possible to consider the boundary condition where 
$X_s(r_h)$ is a nonvanishing constant, but in such cases 
the hairy BHs in scalar-tensor theories are subject to 
instabilities of even-parity perturbations in the vicinity of 
the horizon \cite{Minamitsuji:2022mlv,Minamitsuji:2022vbi} 
(see also Refs.~\cite{Kobayashi:2012kh,Kobayashi:2014wsa,Kase:2021mix,Kase:2023kvq} for the general formulation 
of BH perturbations in Horndeski theories).

In vector-GB theories discussed above, Eq.~(\ref{Xrh}) shows that 
$X$ is a nonvanishing constant at $r=r_h$. 
Unlike scalar-tensor theories, however, we have to caution that 
$X$ contains the temporal vector component $A_0$ besides 
the radial component $A_1$. 
Hence the instability argument performed in Refs.~\cite{Minamitsuji:2022mlv,Minamitsuji:2022vbi} 
for scalar-tensor theories cannot be applied to vector-tensor 
theories. To address this issue, we need to derive linear stability 
conditions of even-parity perturbations in the subclass of 
GP theories. In the most general class of GP theories 
the stability of BHs against odd-parity perturbations was 
addressed in Ref.~\cite{Kase:2018voo}, but 
the analysis in the even-parity sector was not done yet. 
We also note that we only considered the case $r_m=10r_h$ in our 
numerical simulations, but there should be  
appropriate solutions to $A_{\mu}$ for other values of $r_m$, too.
It is beyond the scope of this paper to scrutinize 
all the parameter spaces of boundary conditions as well as 
to study linear stabilities of BHs.

\subsection{Corrections to gravitational potentials}

Let us estimate vector field corrections to the metric 
components $f$ and $h$ both around $r=r_h$ and 
at spatial infinity.
In the vicinity of $r=r_h$, we substitute the expanded solutions 
(\ref{A0expan}) and (\ref{A1expan}) into the gravitational 
Eqs.~(\ref{heq0}) and (\ref{feq0}).
We exploit the leading-order solutions (\ref{fh}) 
for computing corrections to $f$ and $h$ 
of order $\alpha^2$. 
On using the relations (\ref{b0re}) and (\ref{b1re}), 
the differential equations for $h$ and $f$, up to the order 
of $\alpha^2$, are given by 
\ba
& &
-\frac{\Mpl^2}{r^2} \left( rh'+h-1 \right)-\mu_h \alpha^2=0\,,
\label{heq}\\
& &
\frac{\Mpl^2}{r^2} \left( h-1+\frac{rhf'}{f} \right)+\mu_h \alpha^2=0\,,
\label{feq}
\ea
where $\mu_h$ is a constant defined by 
\be
\mu_h \equiv \frac{a_1^2}{2r_h^2}-\frac{a_0 a_1}{r_m^2}
-\frac{4b_0[3a_0 a_1-(2a_0+a_1)b_0+b_0^2]}
{(a_1-b_0)^2 r_h^3}\,.
\ee
We have ignored the corrections of order ${\cal O}(r-r_h)$ 
for the derivation of $\mu_h$.

Now, we search for solutions of the forms 
\ba
f &=& \left( 1-\frac{r_h}{r} \right) \left[ 1+\alpha^2 F(r) \right]\,,
\label{fex} \\
h &=& \left( 1-\frac{r_h}{r} \right) \left[ 1+\alpha^2 H(r) \right]\,,
\label{hex}
\ea
where $F$ and $H$ are functions of $r$. 
Substituting Eq.~(\ref{hex}) into Eq.~(\ref{heq}), we obtain the 
integrated solution 
\be
H(r)=\frac{1}{r-r_h} \left( -\frac{\mu_h r^3}{3\Mpl^2}+{\cal C}_1 
\right)\,.
\ee
The integration constant ${\cal C}_1$ should be chosen to 
avoid the divergence of $H(r)$ at $r=r_h$, such that 
${\cal C}_1=\mu_h r_h^3/(3\Mpl^2)$. 
Then, the solution to $h$ up to the order of $\alpha^2$  
is given by 
\be
h = \left( 1-\frac{r_h}{r} \right) \left[ 1-\alpha^2
\frac{\mu_h (r^2+r_h r+r_h^2)}{3\Mpl^2} \right]
\qquad
{\rm for} \quad r-r_h \ll r_h\,.
\label{hho}
\ee
In the limit that $r \to r_h$, the $\alpha^2$-correction 
in the square bracket of Eq.~(\ref{hho}) approaches 
a constant value $-\alpha^2 \mu_h r_h^2/\Mpl^2$. 
Integrating Eq.~(\ref{feq}) after the substitution 
of Eq.~(\ref{hho}), we obtain 
\be
F(r) =-\frac{\mu_h r (r+r_h)}{3\Mpl^2}+{\cal C}_2\,.
\ee
Setting ${\cal C}_2=0$ by a suitable time reparametrization, 
it follows that 
\be
f = \left( 1-\frac{r_h}{r} \right) \left[ 1-\alpha^2
\frac{\mu_h r (r+r_h)}{3\Mpl^2} \right] 
\qquad 
{\rm for} \quad r-r_h \ll r_h\,,
\label{fho}
\ee
whose $\alpha^2$-order correction is finite around $r=r_h$. 
The fact that the finite corrections to $f$ and $h$ 
arise at the order of $\alpha^2$ is analogous to 
the case of linearly coupled scalar-GB theory. 
We recall however that $\tilde{A}_1$ 
is divergent on the horizon in vector-GB theory, 
while this is not the case for the field derivative 
$\phi'$ in scalar-GB theory.

At large distances, $\tilde{A}_0$ decreases much faster than 
$\tilde{A}_1$. Then, we can ignore contributions of
the $\tilde{A}_0$-dependent terms to the equations 
of motion of $f$ and $h$. 
On using the leading-order solution (\ref{fh}) for the 
computation of $\alpha^2$-order corrections 
arising from the vector field, 
we obtain the following differential equations
\ba
& &
-\frac{\Mpl^2}{r^2} \left( rh'+h-1 \right)
+\frac{192r_h^3 r_m^2}{r^9}\alpha^2=0\,,
\label{heq2}\\
& &
\frac{\Mpl^2}{r^2} \left( h-1+\frac{rhf'}{f} \right)
-\frac{32r_h^3 r_m^2}{r^9}\alpha^2=0\,.
\label{feq2}
\ea
We substitute Eqs.~(\ref{fex})-(\ref{hex}) and their 
$r$ derivatives into Eqs.~(\ref{heq2}) and (\ref{feq2}) 
to obtain the differential equations for $H(r)$ and $F(r)$.
The integrated solution to Eq.~(\ref{heq2}), which is derived by 
setting the integration constant 0 (whose contribution 
can be absorbed into $r_h$), is given by 
\be
h = \left( 1-\frac{r_h}{r} \right) \left( 1-\alpha^2
\frac{32 r_h^3 r_m^2}{\Mpl^2 r^7} \right)
\qquad
{\rm for} \quad r \gg r_h\,.
\label{hho2}
\ee
Thus, the $\alpha^2$-correction term rapidly decreases 
at large distances. On using Eq.~(\ref{hho2}) 
for Eq.~(\ref{feq2}) with Eq.~(\ref{fex}), 
we obtain the integrated solution to $F(r)$. 
Setting the integration constant 0, the resulting 
solution to $f$ is
\be
f = \left( 1-\frac{r_h}{r} \right) \left( 1-\alpha^2
\frac{64 r_h^3 r_m^2}{7\Mpl^2 r^7} \right)
\qquad
{\rm for} \quad r \gg r_h\,.
\label{fho2}
\ee
The $\alpha^2$-order corrections to $f$ and $h$ decrease much faster 
in comparison to linearly coupled scalar-GB theory where 
$F(r)$ and $H(r)$ are proportional to $1/r$ 
at large distances \cite{Sotiriou:2013qea,Sotiriou:2014pfa,Minamitsuji:2022mlv}.

More precisely, when one chooses a specific integration constant to integrate the scalar field equation in scalar-GB theory, the integrated equations of motion are the same as 
Eqs.~\eqref{heq0}-\eqref{A1eq} with $A_0 =0$ and replacing $A_1$ 
with $\phi'$. In scalar-GB theory, the integration constant $Q$ arising from 
the scalar field equation is uniquely fixed to a nonzero 
value \cite{Sotiriou:2013qea, Sotiriou:2014pfa} to realize a finite 
value of $\phi'(r)$ on the horizon \cite{Minamitsuji:2022mlv,Minamitsuji:2022vbi}.  
In vector-tensor theory, the existence of $A_0$ besides $A_1$ allows us 
to satisfy the field equations around the horizon even with a divergent 
value of $A_1$. In this case, two boundary conditions of $A_0$ 
and $A_1$ at $r=r_h$ are not uniquely fixed in general.
Since $A_0$ and the field strength 
$F= -F^{\mu \nu}F_{\mu \nu}/4=hA_0'{}^2/(2f)$ exponentially 
decrease in the regime $r \gg r_h$, the large-distance solution 
is dominated by the longitudinal mode $A_1$ proportional to 
$r^{-5}$. In scalar-tensor theory, the field derivative has the 
large-distance behavior $\phi'(r) \propto r^{-2}$ and hence
its radial dependence is different from that of $A_1$.
In the vicinity of the horizon, the BH solution in vector-GB theory is 
particularly different from that in scalar-tensor theory 
due to the interplay of both temporal and longitudinal 
vector components.

\subsection{BH solutions with $\eta=0$}

Finally, we consider the massless vector field case, i.e., 
\be
\eta=0\,.
\ee
In this case, we can solve Eq.~(\ref{A1eq}) 
for the radial vector component $A_1$ as 
\be
A_1^2=\frac{A_0[(A_0 f'+A_0' f)h+fA_0'-A_0 f'
\pm \sqrt{A_0'f \{A_0' f(h+1)^2 +4A_0f' h (h-1)\}}]}
{f' f h(h-1)}\,.
\label{A1re}
\ee
Around the BH horizon characterized by the radius $r_h$, 
we expand $f$, $h$, and $A_0$ in the forms
\be
f=\sum_{i=1}f_i \left( \frac{r-r_h}{r_h} \right)^i\,,\qquad
h=\sum_{i=1}h_i \left( \frac{r-r_h}{r_h} \right)^i\,,\qquad
A_0=\sum_{i=0}a_i \left( \frac{r-r_h}{r_h} \right)^i\,.
\label{fhA0}
\ee
Substituting Eq.~(\ref{fhA0}) with Eq.~(\ref{A1re}) 
into the background Eqs.~(\ref{heq0})-(\ref{A0eq}), 
the coefficients $f_i$, $h_i$, and $a_i$ can be obtained
at each order.
On using the relation (\ref{A1re}), we have 
\be
X=\frac{a_0 [a_1 \pm \sqrt{a_1 (a_1-4a_0 h_1)}]}{2f_1}
+{\cal O}\left( \frac{r-r_h}{r_h} \right)\,,
\ee
which is finite on the horizon. 
At spatial infinity, one can show that there are 
also solutions respecting the asymptotic flatness. 

Even if there were BH solutions connecting the solutions 
around $r=r_h$ and $r \gg r_h$, the absence of a mass term 
$\eta X$ may induce some instabilities of perturbations 
on the static and spherically symmetric background.
To address this issue, we study the BH stability 
against odd-parity perturbations by using 
linear stability conditions derived 
in Ref.~\cite{Kase:2018voo}. 
For the action (\ref{action}) with $\eta=0$, 
we compute the quantity $q_2$ defined in Eq.~(3.24) 
of Ref.~\cite{Kase:2018voo}.  
Exploiting Eq.~(\ref{A1re}) togehter with 
the expansions (\ref{fhA0}), it follows that 
\be
q_2=-\frac{16 h_1a_0^2 \alpha^2}
{r_h(\Mpl^2 r_h f_1+8 a_0 \sqrt{f_1 h_1}\alpha)
[\sqrt{a_1 (a_1-4a_0 h_1)} \pm a_1]^2 (r-r_h)^2}
+{\cal O} \left( \frac{r_h}{r-r_h} \right)\,. 
\ee
For small $\alpha$, the leading-order term of 
$q_2$ is given by 
\be
q_2=-\frac{16a_0^2 \alpha^2}
{r_h^2\Mpl^2
[\sqrt{a_1 (a_1-4a_0 h_1)} \pm a_1]^2 (r-r_h)^2}\,,
\label{q2}
\ee
where we used the fact that $f_1$ is equivalent to $h_1$ 
in the limit that $\alpha \to 0$. 
The absence of ghosts for vector field perturbations 
requires that $q_2>0$, but we have $q_2<0$ from Eq.~(\ref{q2}). 
Unless $\alpha$ is strictly 0, there is ghost instability 
for small values of $\alpha$.
This problem arises only for the theories with 
$\eta=0$.\footnote{The existence of a problem in the theory with $\eta=0$ can be also seen in the absence of a kinetic term of the longitudinal mode in the decoupling limit $g_{\rm v}\to 0$. 
The longitudinal mode, which is a part of even-parity perturbations in the context of BH perturbations, would be pathological at least in the 
asymptotic region.}

\section{Conclusions}
\label{consec}

We constructed a class of vector-tensor theories in which 
a vector field $A_{\mu}$ is coupled to the vector 
${\cal J}^{\mu}[A, g]$ whose divergence corresponds to 
the GB term ${\cal G}$. 
We showed that the interacting Lagrangian 
$\alpha A_{\mu} {\cal J}^{\mu}$ is equivalent to  
a subclass of GP theories with the quintic coupling 
function $G_5=4\alpha \ln |X|$, where $X=-(1/2)A_{\mu}A^{\mu}$.
This is analogous to the fact that a linear 
scalar-GB coupling of the form $-\alpha \phi \mG$ falls into 
a subclass of Horndeski theories with the coupling 
function $G_{5s}=4\alpha \ln |X_s|$, where 
$X_s=-(1/2)\nabla_{\mu}\phi \nabla^{\mu}\phi$. 
We also extended the analysis to a more general Lagrangian 
$f(X) A_{\mu} {\cal J}^{\mu}$ and found that this belongs to 
a subclass of beyond GP theories given by the action (\ref{SGBf}). 
Since beyond GP theories are the healthy extension of GP 
theories without modifying the dynamical DOFs, 
our vector-GB theories are free from Ostrogradski-type instabilities.

Even though the Lagrangian $\alpha A_{\mu} {\cal J}^{\mu}$ has 
correspondence with the scalar-GB coupling $-\alpha \phi \mG$, 
the fact that the vector field $A_{\mu}$ has a longitudinal
component besides transverse components generally gives rise to 
spacetime dynamics different from those 
in scalar-tensor theories. 
We applied the vector-GB coupling $\alpha A_{\mu} {\cal J}^{\mu}$ 
to the search for static and spherically symmetric BHs 
with vector hairs by incorporating the Einstein-Hilbert term, 
Maxwell scalar, and vector mass term $\eta X$ with $\eta>0$. 
Under an expansion of the small coupling $\alpha$, we derived 
solutions to the temporal and radial vector components both around the 
horizon ($r=r_h$) and at spatial infinity ($r \gg r_h$). 
Unless the boundary conditions on the horizon are chosen 
very accurately, it is difficult to numerically 
integrate the vector field differential equations up to 
sufficiently large distances due to the existence of 
a rapidly growing mode. 
Nevertheless, we confirmed the existence of regular BH 
solutions approaching the large-distance solution for 
some ranges of $r$.

We also computed corrections to the Schwarzschild metric 
arising from the vector field coupled to the GB term. 
At large distances, these corrections rapidly decrease 
in comparison to linear scalar-GB theory. 
On the horizon the radial vector component 
diverges for the coordinate (\ref{spmetric}), 
but this is just a coordinate singularity. In fact, scalar quantities such as $A_{\mu}A^{\mu}$ and $F_{\mu\nu}F^{\mu\nu}$ and the backreaction to the spacetime metric remain finite around $r=r_h$. 
In linear scalar-GB theory, the field kinetic term $X_s$ 
needs to vanish on the horizon 
to avoid linear instability of even-parity 
perturbations. In this case, the perturbatively derived 
BH solution is uniquely determined for a given small 
coupling $\alpha$. 
In our vector-GB theory, we generally have a nonvanishing 
value of $X$ on the horizon. 
For small couplings $\alpha$, we numerically found that 
there are some ranges of boundary conditions of the vector field 
in which analytic solutions in two asymptotic regimes 
are joined each other.
This suggests that, unlike linear scalar-GB theory, hairy 
BH solutions in vector-GB theory for given $\alpha$ 
are not unique. 

The fact that $A^{\mu}$ cannot be expressed in terms of a scalar gradient 
due to the nonvanishing field strength $F=hA_0'{}^2/(2f)$ 
differentiates our BH solution from that in scalar-GB theories. 
At large distances, $A_0$ decays rapidly relative to the 
longitudinal mode $A_1$. The latter asymptotic solution is given by  
$A_1 \propto r^{-5}$, whose radial dependence is different 
from a large-distance solution of the field derivative 
($\phi' \propto r^{-2}$) in scalar-GB theory.
In the vicinity of the horizon, the contribution of $A_0$ 
to the vector-field equation of motion is as important as that of 
$A_1$. Thus, the boundary conditions on the horizon are different 
from those in scalar-tensor theories. 
This means that taking the scalar limit $A^{\mu} \to \nabla^{\mu} \phi$  
for our BH solution
does not recover the hairy BH in scalar-GB theory.

It will be of interest to search for the parameter space 
of boundary conditions in more detail for broader ranges of
the vector field mass. We would also like to 
formulate BH perturbations in GP theories especially 
for the even-parity sector to study the linear stability of 
hairy BHs. As a byproduct, it will be possible to compute the 
quasi-normal modes of BHs which can be probed by the gravitational 
wave observations. These issues are left for future works.

\section*{Acknowledgements}

The work of K.A. was supported in part by Grants-in-Aid from the Scientific Research Fund of the Japan Society for the Promotion of Science, No.~20K14468.
S.T. was supported by the Grant-in-Aid for Scientific Research 
Fund of the JSPS Nos.~19K03854 and 22K03642.

\bibliographystyle{mybibstyle}
\bibliography{bib}

\end{document}